\documentstyle[aps,multicol]{revtex}
\begin{document}
\draft
\title{Gauge Invariance and Finite Temperature Effective Actions of
Chern-Simons Gauge Theories with Fermions}
\author{Daniel Cabra
$^a$,  Eduardo Fradkin $^b$,
Gerardo L.Rossini $^a$ \\
and Fidel A.Schaposnik $^a$\\
{\normalsize \it Departamento de
F\'\i sica, Universidad Nacional de La Plata}$^a$ \\
{\normalsize \it C.C. 67, (1900) La Plata, Argentina.}\\
{\normalsize \it and }\\
{\normalsize \it Department of Physics, University of Illinois at
Urbana-Champaign}$^b$\\
{\normalsize \it 1110 W.~Green St.~, Urbana, IL 61801, USA}\\}
%
\maketitle
\begin{abstract}
{We discuss the behavior of theories of fermions coupled to Chern-Simons gauge
fields
with a non-abelian gauge group
in three dimensions  and at finite temperature. Using non-perturbative
arguments and
gauge invariance, and in contradiction with perturbative results, we show
that the
coefficient of the Chern-Simons term of the  effective actions for the gauge
fields
at finite temperature can be {\it at most} an integer function of the
temperature.
This is in  a sense a  generalized no-renormalization theorem.
We also discuss the case of abelian theories  and give  indications that a
similar
condition should hold there too.  We discuss consequences of our results to
the thermodynamics of anyon superfluids and fractional quantum Hall systems.}
\end{abstract}
\bigskip

\pacs{PACS numbers: 11.10.Kk,11.10.Wx,11.15.-q,11.30.Er, 73.40.Hm,74.20.Kk}

\begin{multicols}{2}

\columnseprule 0pt

\narrowtext

\newpage

In the past decade a significant amount of effort has been devoted to study the
behavior of field theories of matter coupled to Chern-Simons gauge fields. At
zero temperature, pure Chern-Simons gauge theories are topological field
theories which by now are well understood\cite{wittenjones}. Much less is known
about theories in which the gauge fields are coupled to matter. For the most
part, these theories have been treated in perturbation theory or in cases in
which the gauge fields are non-dynamical background fields. These results have
shown that the fluctuations of a massive Fermi field induces a Chern-Simons
term in the effective action of the gauge fields. This is the celebrated Parity
Anomaly of Deser, Jackiw and Templeton\cite{DJT}.
 The result of these studies is that the Chern-Simons coupling constant (often
referred to as the topological mass) is equal to $\theta/4 \pi$, where the
parameter $\theta$ is an integer equal to the number of fermion
species\cite{schonfeld,DJT}. It was also found\cite{colemanhill} that this
one-loop result does not get renormalized by higher order loop corrections.
For non-abelian pure Chern-Simons theories topological arguments have
shown\cite{DJT} that for a theory on a compact manifold the coefficient of the
Chern-Simons action must be quantized, both at the classical and at the quantum
level.

Non-relativistic versions of these theories have also been investigated.  It
has been established that these theories have superfluid ground states and have
given a concrete model of anyon superconductivity. It was found that the
superfluidity of the ground state is ensured by the exact cancellation of the
coefficient of the one-loop induced Chern-Simons term against the bare
Chern-Simons coupling constant which sets the fractional
statistics\cite{banks,efanyons}.
Theories of non-relativistic matter coupled to Chern-Simons fields are commonly
used in the study of the physics of the two dimensional electron gas in strong
magnetic fields and provide a natural theoretical framework for the Fractional
Quantum Hall effect\cite{zhk,wen,lopez}. Here too, the non-renormalization of
the induced Chern-Simons terms in the effective action of the gauge fields is
fundamental for the theory to be consistent with the requirements of Galilean
invariance and for the Hall conductance to be determined by the filling
fraction of the system.

Much less is understood at finite temperatures. Perturbative  calculations, for
both relativistic and non-relativistic theories, abelian and non-abelian, have
in almost all cases, yielded induced actions with Chern-Simons coefficients
which are smooth functions of the temperature\cite{NS}-\cite{F}. However,
Pisarski \cite{Pis} has argued that the exact answer should be a constant,
independent of the temperature. It should be emphasized that, in all these
calculations, the Chern-Simons gauge fields were taken to be non-dynamical
background fields.

There are significant physical reasons to suspect that an induced action with
Chern-Simons coefficients which are smooth functions of the temperature cannot
possibly be the right answer for the full theory. For non-abelian theories it
is hard to believe that the topological arguments that lead to the exact
quantization of the Chern-Simons coefficient at zero temperature could not be
extended to finite temperature (even though the manifold is no longer a
sphere). For abelian theories, the exact cancellation between the induced and
bare Chern-Simons terms, required for anyon superfluidity to work, would be
violated at finite temperature if this results would hold literally. In fact,
several authors\cite{weiss,hosotani} have advocated a picture in which anyon
superfluidity sort of ``evaporates" at any non-zero temperature. A priori, on
general grounds, one expects that an anyon superfluid should undergo a
Kosterlitz-Thouless type transition at a non-zero critical temperature rather
than an immediate destruction for all $T >0$. A loophole in the perturbative
approach is hinted by the following argument.
The fermion excitations which renormalize the induced Chern-Simons coefficient
in a temperature dependent fashion are gauge non-invariant states which  cannot
be part of the physical spectrum and, as such, they should not be included in
the partition function. Furthermore, in the case of the anyon superfluid, these
states have a logarithmically divergent self-energy.  Thus, their weight in the
partition function should vanish. Naturally, fermion-antifermion pairs are
allowed finite energy excitations whose energy grows logarithmically with the
pair size. In a sense, the fermion should be viewed as a vortex (as in the
Kosterlitz-Thouless theory) and the unbinding of these pairs should trigger the
actual phase transitions. However, the removal of these gauge non-invariant
states from the partition function can only be achieved by including explicitly
the fluctuations of the gauge fields and it is a non-perturbative effect.
Thus, the mechanism which makes sure that the Chern-Simons coefficient is an
integer is also responsible for a finite temperature phase transition.

In this paper we reexamine the properties of the effective action of the gauge
fields at finite temperature and its consistency with the requirement of gauge
invariance. We show that effective actions with smooth, temperature-dependent
renormalizations of the Chern-Simons coupling constant are inconsistent with
gauge invariance. We show that these coupling constants can be, at most,
integer functions of the temperature. For the case of non-abelian theories, we
show that the presence  of configurations of gauge transformations with
non-trivial winding number forces the quantization of the coefficients of the
induced and bare Chern-Simons terms {\it at all temperatures}. Although, as
they stand, our results apply directly only to non-abelian theories, we claim
that the quantization of the induced term at finite temperature should apply
for abelian theories as well.

We start from the three-dimensional (Euclidean) action
\begin{equation}
S =  \int d^3x \bar\psi
(i\kern.06em\hbox{\raise.25ex\hbox{$/$}\kern-.60em$\partial$}
+ m + \not\!\! A) \psi +
\frac{\theta}{4\pi} S_{CS}[A]
\label{a}
\end{equation}
where $S_{CS}[A]$ is the Chern-Simons action
\begin{equation}
  S_{CS}[A] =
 i \epsilon_{\mu\nu\lambda} {\rm tr} \int d^3x
  (
   F_{\mu \nu} A_{\lambda} -
   \frac{2}{3} A_{\mu}A_{\nu}A_{\lambda}
  )  .
\end{equation}
for a gauge field $A_\mu$ taking values in the Lie algebra of
some gauge group $G$,
\begin{equation}
F_{\mu\nu} =
  \partial_{\mu}A_{\nu} -
  \partial_{\nu}A_{\mu} +
  [A_{\mu},A_{\nu}]  .
\label{curv}
\end{equation}
As it is well known, the non-Abelian Chern-Simons action,
 is
not gauge invariant but changes under large gauge transformations
\cite{DJT},

\begin{equation}
A_\mu \to  A^g_\mu = g^{-1} A_\mu g + g^{-1} \partial g
\label{AAA}
\end{equation}
\begin{equation}
S_{CS}[A]  \to  S_{CS}[A^g] = S_{CS}[A] + 8\pi^2 i w[g]
\label{b}
\end{equation}
\begin{equation}
w[g] = \frac{1}{24\pi^2} \int d^3x \epsilon^{\mu \nu \alpha}
g^{-1} \partial_{\mu} g g^{-1} \partial_\nu g g^{-1} \partial_\alpha g
\label{w}
\end{equation}

Finite temperature calculations are carried out as usual by
compactifying the (Euclidean) time variable into the range
$0 \le \tau \le \beta = 1/T$ (in our units, $\hbar = c = k = 1$).
Concerning bosonic (fermionic) fields, periodic (antiperiodic)
boundary conditions (in time) have to be used. Moreover,
we shall assume that the allowed gauge transformations are
periodic in time. Compactifying space-time, the resulting
manifold is $S^2 \times S^1$ and one can then prove \cite{Al} that
for compact gauge groups, $w[g]$ is the winding number of $g$,
$w[g] \in Z $
. In writing eq.(\ref{b})
we have dropped a surface term which vanishes
provided $A_\mu$ vanishes rapidly enough at spatial
infinity.
{}From this analysis, we see that
$\theta$ should be chosen to be an integer if $\exp (-S)$ is to be gauge
invariant also for large gauge transformations \cite{DJT}.

The partition function at finite temperature is then defined
as
\begin{equation}
{\cal Z} =
  {\cal N}(\beta) \int {\cal D}\bar{\psi} {\cal D}\psi DA_\mu \>
                  \exp\left( -S_\beta \right) ,
\label{Z}
\end{equation}
where $S_\beta$ is the action (\ref{a}) at finite temperature,
${\cal N}(\beta)$ is a temperature dependent normalization
constant and, as stated above, one integrates over gauge fields
and fermion fields satisfying periodic and antiperiodic (in time)
boundary conditions respectively.
Integrating the fermions out we are left with
\begin{equation}
{\cal Z} =
  {\cal N}(\beta) \int DA_\mu \> \exp (-\frac{\theta}{4\pi} S_{CS}[A]) \times
\det (i\kern.06em\hbox{\raise.25ex\hbox{$/$}\kern-.60em$\partial$} + m +
\not\!\! A)
\label{Z1}
\end{equation}

It is important to stress that the gauge field integration in (\ref{Z1})
ranges over all (periodic) gauge field configurations. One can make this
explicit by means of the Faddeev-Popov procedure \cite{FP} which consists
in writing the measure $DA_\mu$ in the form
\begin{equation}
DA_\mu = DA_\mu \delta[F[A^g]] \Delta_{FP}[A] Dg
\label{FP}
\end{equation}
where $\Delta_{FP}$ is the Faddeev-Popov determinant associated
with the $F[A] = 0$ condition and $Dg$ is the usual Haar measure
giving the volume element on the group of gauge transformations.
After inserting (\ref{FP}) in (\ref{Z1}) and changing $A^g\to A$,
$g \to g^{-1}$,
one gets
\begin{eqnarray}
{\cal Z} & = &
  {\cal N}(\beta) \int DA_\mu \delta[F[A]] \Delta_{FP}[A] Dg
\nonumber \\
& & \exp (-\frac{\theta}{4\pi} S_{CS}[A^{g}]) \times \det
 (i\kern.06em\hbox{\raise.25ex\hbox{$/$}\kern-.60em$\partial$} + m +
\not\!\! A ^g)
\label{r}
\end{eqnarray}
It will be useful to define an effective action $S_{eff}[A]$
in the form
\begin{eqnarray}
\exp (-S_{eff}[A]) &=& \nonumber \\
\int Dg \;
\exp (-&&\!\! \frac{\theta}{4\pi} S_{CS}[A^{g}]) \times \det
(i\kern.06em\hbox{\raise.25ex\hbox{$/$}\kern-.60em$\partial$} + m +
\not\!\! A ^g)
\label{ef}
\end{eqnarray}
so that
\begin{equation}
{\cal Z}  =
  {\cal N}(\beta) \int DA_\mu \delta[F[A]] \Delta_{FP}[A]
\exp (-S_{eff}[A])
\label{ef1}
\end{equation}
If the Chern-Simons action and the fermion determinant were gauge invariant,
the group integration in eq.~(\ref{ef}) would be trivial and the effective
action would be the one that is usually expected.

One can easily check that the effective action (\ref{ef})
is gauge invariant. Indeed,
\begin{eqnarray}
\exp (-S_{eff}[A^h])  &=& \nonumber \\
\int Dg \;
\exp (-&&\!\!\frac{\theta }{4\pi}S_{CS}[A^{h g}])
\times \det (i\kern.06em\hbox{\raise.25ex\hbox{$/$}\kern-.60em$\partial$} + m +
\not\!\! A ^ {h g})
\label{effi}
\end{eqnarray}
or, after making the change of variables $g \to h^{-1}g$
\begin{equation}
\exp (-S_{eff}[A^h]) = \exp (-S_{eff}[A])
\label{coc}
\end{equation}
This, of course, follows from the fact that the (properly regularized)
fermion determinant  satisfies
a natural condition of consistency with the group property of
the gauge transformations. Indeed, defining
\begin{equation}
\exp(i \alpha[A,g]) = \det
(i\kern.06em\hbox{\raise.25ex\hbox{$/$}\kern-.60em$\partial$} + m +
\not\!\! A ^ { g})/
\det (i\kern.06em\hbox{\raise.25ex\hbox{$/$}\kern-.60em$\partial$} + m +
 \not\!\! A )
\label{coci}
\end{equation}
one has the $1$-cocycle condition
\begin{equation}
\delta \alpha = \alpha[A^g,h] - \alpha[A,g h] + \alpha[A,g] = 0
\label{fin}
\end{equation}
This property is at the root of consistent quantization
of anomalous gauge theories since it ensures the
gauge invariance of the effective action \cite{anom}-\cite{ht}.
Moreover, similar arguments about
gauge invariance can be used for proving
that monopole contributions are wiped out, at zero temperature,
from the $QED_3$ generating functional \cite{Aff}-\cite{FS}.

The arguments presented above ensure also in the
present case that the {\it exact} effective
action is gauge invariant. The exact expression for
the fermion determinant in three space-time dimensions is not known. However,
there is an extensive
literature on the approximate form of this determinant at finite
temperature. Perturbative arguments suggest that the anomalous
part of the fermion determinant has the form \cite{NS}-\cite{F}
\begin{equation}
\det (i\kern.06em\hbox{\raise.25ex\hbox{$/$}\kern-.60em$\partial$} + m +
\not\!\! A) =
\exp (-\frac{1}{4\pi}F[T] S_{CS}[A])
\label{aqui}
\end{equation}
Almost without exception, the results on this subject,
based in perturbation theory,
yield a function  $F[T]$ which is a function of the temperature
\begin{equation}
F[T] = \frac{1}{2} \tanh ( \frac{m\beta}{2} )
\label{T}
\end{equation}
This result holds for both abelian and non-abelian theories.
In contrast, Pisarski \cite{Pis}, using a non-perturbative formal
argument  claimed that, for a non-abelian theory, this coefficient is actually
temperature independent and equal to its value at zero temperature.

We shall now discuss the consistency of the ansatz (\ref{aqui})
with the requirement of gauge invariance.
Thus, we will assume that eq.(\ref{aqui}) holds and write
\begin{equation}
\exp (-S_{eff}[A]) = \int Dg
\exp \left (-\frac{1}{4\pi}(\theta + F[T]) S_{CS}[A^{g}]\right)
\label{ef11}
\end{equation}
or, after (\ref{b})
\begin{eqnarray}
\exp (-S_{eff}[A]) &=& \exp \left(
-\frac{1}{4\pi}(\theta + F[T])
S_{CS}[A]\right) \times \nonumber \\ \
& & \int Dg \exp
 \left(- 2\pi i (\theta +  F[T]) w[g] \right)
\label{q}
\end{eqnarray}
We can now split up the integral over $g$ in sectors $g^{(n)}$
according to $w[g^{(n)}] = n$
\begin{eqnarray}
\exp (-S_{eff}[A]) & = &\exp \left(
-\frac{1}{4\pi}(\theta + F[T])
S_{CS}[A]\right) \times \nonumber \\
& &
\sum_{n = - \infty}^{\infty} \int Dg^{(n)} \exp
 \left(2\pi i n(\theta +  F[T]) \right)
\label{qz}
\end{eqnarray}
One can easily see that the Haar measure $Dg^{(n)}$
is the same for all topological sectors, labelled by the winding number  $n$,
so that the
integral for each $n$  gives the same factor (the volume of the gauge
group $V[G]$). Hence one ends up with
\begin{eqnarray}
\exp (-S_{eff}[A]) &=& V[G] \exp \left(
-\frac{1}{4\pi}(\theta + F[T])
S_{CS}[A]\right) \times \nonumber \\
& & \sum_{n = - \infty}^{\infty}\exp\left(2\pi i n(\theta + F[T]) \right)
\label{qq}
\end{eqnarray}
We recognize in the last factor a representation of the
(periodic) delta function
\begin{eqnarray}
\exp (-S_{eff}[A]) &=& \exp \left(
-\frac{1}{4\pi}(\theta + F[T])
S_{CS}[A]\right)
\nonumber \\ &\times&
\sum_{k = - \infty}^{\infty}\delta \left( \theta + F[T] - k \right)
\label{qq1}
\end{eqnarray}
Hence, the partition function vanishes unless
the following constraint is satisfied
\begin{equation}
F[T]  + \theta = 0 ~~ [{\rm mod} ~~k]
\label{fff}
\end{equation}
This is the main result in our work. It states that $F[T]$, the
coefficient of the Chern-Simons term induced by the fermion
integration, must be an integer valued function of the
temperature since the model should be consistent even for $\theta = 0$.
This, in turn, implies that in general $\theta $ should be,
as it is the case at $T=0$, an integer.

A similar line of argument can be followed for the case of an {\it abelian}
gauge group. However, in this case there is no topological invariant
analogous to $w[g]$.
Thus, this line of reasoning does not yield any directly useful information for
abelian theories. However, as we will see below, the non-abelian result has
potentially far reaching implications even for the abelian case.

It should be noted that perturbative approaches leading to the temperature
dependence described by eq.(\ref{T}) (both in the Abelian and
non-Abelian cases) are all based on a $\not\!\! p/m$ expansion.
Now, in the $m \to \infty$ limit, the coefficent of the Chern-Simons
term as given by (\ref{T}) tends to a step function,
up to exponentially small corrections, which precisely
satisfies the necessary condition (\ref{fff}).

A number of important conclusions can be drawn from our results.
They imply
 that  the effective action for {\it dynamical} gauge fields
at finite temperature {\it cannot} contain a Chern-Simons term with a
coefficient which is a smooth function of the temperature.
For the case of non-abelian theories the Chern-Simons coefficient
is quantized at zero temperature.
Our results show that it must also be an integer at any non-zero
temperature. Notice that our argument does not exclude the possibility
of a temperature dependence  of this coefficient.
It states that {\it at most} it must be an integer-valued
function of the temperature. Consequently, it can change only by integers
at different ranges of temperature.
Thus, up to integer shifts, the Chern-Simons coupling constant remains
unrenormalized. Therefore we conclude that the perturbative approaches
which yield Chern-Simons coefficients with a smooth temperature
dependence, are {\it inconsistent} with the requirement of gauge
invariance.
Since the same perturbative approaches yield the same result
 also for  abelian theories, we strongly suspect that they must also fail
in the abelian case in spite of the fact that our arguments do not give any
useful information for abelian theories. Thus our results give a strong
hint that, quite generally, the Chern-Simons coefficient does not get
 smoothly renormalized at non-zero temperature.

DC and GLR are members of CONICET (Argentina) and FAS is an
Investigador CICBA (Argentina).
This work was
supported in part by the National Science Foundation through the
grant NSF DMR-94-24511 at the University of Illinois at Urbana
Champaign (EF), by CICBA and CONICET (FAS), by the NSF-CONICET
International Cooperation Program through the grant NSF-INT-8902032
and by Fundaci\'on Antorchas through the grant A-13218/1-153.
F.~S.~ would like to thank C.~Fosco for useful comments.
E.~F.~ thanks the Universidad de La Plata for its kind hospitality.

%

\end{multicols}

\end{document}